\documentclass[twoside,12pt]{article}
\usepackage{amsmath,amsfonts,amssymb}
\usepackage[usenames,dvipsnames,svgnames,table]{xcolor}
\usepackage{graphicx}
\usepackage{subfigure}
\newcommand{\nc}{\newcommand}
\nc{\la}{\lambda} \nc{\alf}{\alpha} \nc{\La}{\Lambda}
\nc{\tht}{\theta}  \nc{\T}{\Theta}  \nc{\be}{\beta}  \nc{\eps}{\epsilon} 
\nc{\ga}{\gamma}  \nc{\De}{\Delta}  \nc{\G}{\Gamma}  \nc{\vphi}{\varphi}

\nc{\de}{\delta} \nc{\si}{\sigma}  \nc{\ka}{\kappa}   \nc{\Si}{\Sigma} 
\nc{\om}{\omega}  \nc{\qq}{\quad\quad}                \nc{\Om}{\Omega}
\nc{\nf}{\infty}   \nc{\dl}{\mathop{\smash{\cal L}}}  \nc{\black}{\rule{3mm}{3mm}}
\nc{\ra}{\rightarrow}    \nc{\ol}{\overline}        \nc{\und}{\underline} 
\nc{\beq}{\begin{equation}}  \nc{\eeq}{\end{equation}}  \nc{\pt}{\partial}  
   \nc{\dst}{\displaystyle}  \nc{\na}{\nabla} 
\nc{\nnb}{\nonumber}    \nc{\bs}{\backslash}        \nc{\mb}{\mathbb}   
\nc{\sn}{{\rm sn}\,} \nc{\cn}{{\rm cn}\,}     \nc{\dn}{{\rm dn}\,}
\nc{\ti}{\tilde}   \nc{\wti}{\widetilde}   \nc{\h}{\hat}  \nc{\wh}{\widehat}   \nc{\nin}{\noindent}
\nc{\tpsi}{\wti{\psi}}   \nc{\tphi}{\wti{\phi}}  \nc{\tH}{\wti{H}} \nc{\Ai}{{\rm Ai}}

\newcounter{muni}
\newenvironment{remunerate}{\begin{list}{{\rm \arabic{muni}.}}
{\usecounter{muni}
\setlength{\leftmargin}{0pt}\setlength{\itemindent}{38pt}}}{\end{list}}

\nc{\brm}{\begin{remunerate}}   \nc{\erm}{\end{remunerate}}

\newtheorem{nth}{Theorem}
\nc{\stg}{\mathop{\smash{*}}}
\nc{\st}{\mathop{\smash{\delta}}}
\nc{\stL}{\mathop{\smash{\cal L}}}
\nc{\barr}{\begin{array}}   \nc{\earr}{\end{array}}   \nc{\dg}{\dagger}
\nc{\mtvb}{\mathversion{bold}}   \nc{\mtvn}{\mathversion{normal}}

\title{\bf The geodesic flow of the BGPP metric \\ is  Liouville integrable}

\author{
	Andrzej J. Maciejewski \\
	Janusz Gil Institute of Astronomy, University of Zielona G\'ora, \\
	ul. Licealna~9, 65-417,  Zielona G\'ora, Poland \\
	e-mail: a.maciejewski@ia.uz.zgora.pl \\ 
	Maria Przybylska \\
	Institute of Physics, University of Zielona G\'ora, \\
	Licealna 9, 65--417,  Zielona G\'ora, Poland
	\\ e-mail: m.przybylska@if.uz.zgora.pl
 \\ and \\
 Galliano Valent\\
 LPMP: Institut de Physique Mathématique de Provence\\
 13100 Aix en Provence, France\\
 e-mail: galliano.valent@orange.fr
 }

\date{}
\begin{document}
\maketitle

\begin{abstract}
We prove that the geodesics equations corresponding to the  BGPP metric are
integrable in the Liouville sense.  The $\mathrm{SO}(3,\mathbb{R})$ symmetry of
the model allows to reduce the system from four to two degrees of freedom.
Moreover, solutions of the reduced system and its degenerations  can be solved
explicitly or reduced to a certain quadrature.  In degenerated cases BGPP metric
coincides with the Eguchi-Hanson  metric and for this case the mentioned
quadrature can be calculated explicitly in terms of elliptic integrals. 
    
\end{abstract}

\section{Introduction}
When looking for gravitational instantons was more fashionable, the articles
describing new hyperk\"ahler metrics flourished. Among these metrics, two
Euclidean triaxial metrics, with the Lie algebra  $\mathfrak{su}(2)$ of
infinitesimal isometries,  were impressive. The first one, the BGPP metric,  was found by Belinski, Gibbons, Page, and Pope in 1978, see \cite{BGPP}.  The second one, called the Atiyah-Hitchin (A-H) metric, was introduced in \cite{AH1} and
its relation with monopole scattering was developed in \cite{AH2}. 

The BGPP metric is still of importance due to its role of a seed in the
development of geometries in higher dimensions: in \cite{Do} Einstein 5D
geometries with a negative cosmological constant, in \cite{Gh}  Einstein-Maxwell
metrics for $D\geq 6$ and in \cite{KL} a supersymmetric black-hole in 5D.

Although both these metrics are triaxial, they are somewhat different in
complexity. Indeed, components of the  BGPP  metric are given by simple
algebraic functions, while in the case of A-H they are given by elliptic
integrals, and it seems that they cannot be reduced to a simpler form.

A second difference is that the first metric is not \emph{complete} while the
second one, which describes the relative motion of two magnetic monopoles,
happens to be complete.

Both metrics exhibit biaxial limits: Eguchi-Hanson for the BGPP metric and
Taub-NUT for the A-H metric \cite{GM}.

The BGPP metric lies in the multicentre family as proved in \cite{GOVR},
\cite{Gi}, characterized by at least one tri-holomorphic Killing vector, while
A-H metric does not belong to this family.

Considering only the multicentre metrics, the analysis of the integrability of
their geodesic flow was developed in \cite{GR} for Taub-NUT and the two-centre
metrics while in  \cite{Va1} a more general analysis led to seven metrics mostly
related with Bianchi A metrics as shown in \cite{Va2}.

Despite these results and further work in \cite{Gi} and in \cite{Va2}, the
integrability status of the BGPP metric remained unknown. 

The aim of this article is to study the integrability of their geodesic flow. In
Section~\ref{sec:BGPP}  we review the known integrability results for the
Multicentre metrics in order to have a better insight of the BGPP metric. In
Section~\ref{sec:Hamiltonian} we lay down the  Hamiltonian formulation of the geodesic flow of this metric and show the integrability in the Liouville sense
of this flow. In Section~\ref{sec:reduced_system} we analyse the corresponding reduced five-dimensional system which is a Poisson system with one Casimir
function and on a symplectic leaf of this Poisson structure this system is
Hamiltonian with two degrees of freedom and appears to be integrable in the
Liouville sense. Moreover, in Section~\ref{sec:reducedintegration} we describe
its solutions expressible by the Jacobi elliptic functions $\operatorname{sn}$,
$\operatorname{cn}$ and  $\operatorname{dn}$.
Section~\ref{sec:degenerated_system} is devoted to degenerate cases of the 
BGPP metric which appear to coincide with metrics I and II from paper of Eguchi
and Hanson \cite{EH}. The corresponding geodesics flows are integrable as well
as the corresponding reduced systems are integrable and their solutions are
expressible either by elliptic or by trigonometric functions. Some conclusions are presented in
Section~\ref{sec:conclusion}.

\section{Multicentre versus BGPP metric}
\label{sec:BGPP}
The Multicentre metrics are four dimensional Euclidean metrics. Taking for coordinates $(\si,\,x^i), \ i=1,2,3$ one has
\beq\label{MCgen}
g=\frac 1V(d\si+\om)^2+V\,\ga_0,
\eeq
where $V$ and $\om$ are independent of $\si$ and $\ga_0=
d\vec{x}\cdot d\vec{x}$, where $\vec{x}=\left(x^1,x^2,x^3\right)$, is the flat metric in ${\mb R}^3$. It follows that all these metrics share the Killing vector ${\cal K}=\pt_{\si}$. Taking 
\[ e_0=\frac 1{\sqrt{V}}(d\si+\om),\qq\qq e_i=\sqrt{V}\,dx^i,\quad i=1,2,3 \]
one can construct a triplet of two-forms given by
\beq
\Om_i=e_0\wedge e_i+\frac12\eps_{ijk}\,e_j\wedge e_k=\frac 12 (J_i)_{\mu\nu}\,dx^{\mu}\wedge dx^{\nu},\qq i=1,2,3 ,
\eeq
with the quaternionic multiplication table
\[J_i\ J_j=-\de_{ij}\,Id+\eps_{ijk}\,J_k.\]
The closedness of these two-forms follow from
\beq
d\om=\stg_{\ga_0}\,dV \qq\Longrightarrow\qq \De V=0.
\eeq
By a theorem of Hitchin \cite{Hi} this construction produces a family of hyperk\"ahler hence Ricci-flat metrics characterized by tri-holomorphic complex structures:
\beq \stL_{{\cal K}}\,J_i=0,\qq i=1,2,3. \eeq

The quest for integrability relies on the existence of a (symmetric) Killing-St\"ackel (K-S) tensor $S_{\mu\nu}$ such that
\[ \nabla_{(\mu}\,S_{\nu\rho)}=0,\]
which gives a quadratic integral
\[ {\cal S}=S^{\mu\nu}\,P_{\mu}\,P_{\nu},\qq\qq \{H,{\cal S}\}=0.\]

The first results establishing the Liouville integrability of the flow for the Multicentre metrics began with the work of Gibbons and Manton \cite{GM} for the Taub-NUT metric for which
\[ V=1+\frac mr,\]
and was generalized in \cite{GR} to the two-centre metric for which
\[ V=v_0+\frac{m_1}{|\vec{r}+\vec{c}|}+\frac{m_2}{|\vec{r}-\vec{c}|}.\]
In the first case the construction of a generalized Runge-Lenz vector allowed for an elegant geometric derivation of the geodesics while in the second case an educated guess of a set of coordinates separating the Hamilton-Jacobi equation led to the Killing-Stackel tensor. 

A more systematic analysis in \cite{Va1} led to the following result: let it be supposed that $V$ has one Killing vector i.e. $K^i\,\pt_iV=0$. There are 8 choices of coordinates separating the Hamilton-Jacobi in $\ga_0$
(among the 11 possible choices) which lift up to separating coordinates for the Hamilton-Jacobi equation for $g$, producing in each case a quadratic Killing-Stackel tensor. The Liouville integrability stems from the set of integrals
\[ H,\qq P_{\si},\qq K=K^i\,P_i, \qq {\cal S},\]
which include two linear ones and two quadratic ones. This is no longer the case for the BGPP metric. 

The BGPP metric \cite{BGPP} is given by
\beq\label{met0}
g=f(t)^2\,dt^2+a(t)^2\,\si_1^2+b(t)^2\,\si_2^2+c(t)^2\,\si_3^2,\qq 
d\si_i=\frac 12\,\eps_{ijk}\,\si_j\wedge\si_k.
\eeq
where  $\si_i$ are the Maurer-Cartan 1-forms for the group $G=\mathrm{SU}(2)$. Using Euler angles, we have
\[ \barr{lrr} \si_1= & -\sin\psi\,d\tht & +\sin\tht\,\cos\psi\,d\phi,\\
\si_2= & +\cos\psi\,d\tht & +\sin\tht\,\sin\psi\,d\phi,\\
\si_3= & d\psi & +\cos\tht\,d\phi.\earr \]

For the BGPP metric one has 
\beq
f(t)^2=\frac 1{4A(t)B(t)C(t)},\,\,\,\, a(t)^2=\frac{B(t)C(t)}{A(t)}, \,\,\,\, b(t)^2=\frac{C(t)A(t)}{B(t)}, \,\,\, \,c(t)^2=\frac{A(t)B(t)}{C(t)},
\eeq
with
\beq
A(t)=\sqrt{t-t_1},\qq B(t)=\sqrt{t-t_2}, \qq C(t)=\sqrt{t-t_3},\eeq
compare with formulas (12) and (13) in \cite{BGPP}. The real three parameters $t_1,t_2,t_3$ are non-negative and can be chosen arbitrarily. The coordinate $t$ is restricted to  the half line $t>t_m$, where $t_m$ is the biggest parameter $t_i$. 

There is the   algebra  $\mathfrak{g}=\mathfrak{su}(2)$    of infinitesimal isometries 
\beq\barr{l}\dst 
\wh{\cal L}_1=-\sin\phi\,\pt_{\tht}+\frac{\cos\phi}{\sin\tht}\,\pt_{\psi}-\frac{\cos\phi}{\tan\tht}\,\pt_{\phi},
\\[4mm] \dst
\wh{\cal L}_2=-\cos\phi\,\pt_{\tht}-\frac{\sin\phi}{\sin\tht}\,\pt_{\psi}+\frac{\cos\phi}{\tan\tht}\,\pt_{\phi},\\[4mm] 
\wh{\cal L}_3=\pt_{\phi},\earr\qq\qq [\wh{\cal L}_i,\wh{\cal L}_j]=\eps_{ijk}\,\wh{\cal L}_k.
\eeq

Using the co-frames
\[
e_0=f(t)dt,\quad e_1=a(t) \si_1,\quad e_2=b(t) \si_2,\quad e_3=c(t) \si_3,
\]
it is easy to check that the self-dual 2-forms 
\beq 
\Om_i=e_0\wedge e_i+\frac 12\,\eps_{ijk}\,e_j\wedge e_k\qq i=1,2,3
\eeq
are closed since we have
\beq
\Om_1=d(A(t)\,\si_1),\qq \Om_2=d(B(t)\,\si_2),\qq \Om_3=d(C(t)\,\si_3).
\eeq
Since the Killing vector $\wh{\cal L}_3=\pt_{\phi}$ is  tri-holomorphic the BGPP metric belongs to the Multicentre family.
Indeed,  taking $d\si=d\phi$ it was shown in \cite{GOVR} that the potential is given by
\beq
V=\frac{1}{a(t)^2\sin^2\tht\cos^2\psi+b(t)^2\sin^2\tht\sin^2\psi+c(t)^2\cos^2\tht},
\eeq
and one-form $\omega$ is 
\beq
\omega=V\left[\left(b(t)^2-a(t)^2\right) \sin \theta 
   \sin \psi \cos \psi d\theta+c(t)^2\cos
   \theta d\psi
\right].
\eeq
In these settings coordinates  $\vec{x}=\left(x^1,x^2,x^3\right)$ are
\[ 
x^1=A\sin\tht\cos\psi,\qq x^2=B\sin\tht\sin\psi,\qq x^3=C\cos\tht.
\]
The problem is that this potential has {\em no symmetry at all.} 
This explains why, despite the analyses developed in \cite{Va2} and \cite{Gi}, the integrability status of its geodesic flow remained unknown for 45 years...

\section{Hamiltonian formulation of the geodesic flow}
\label{sec:Hamiltonian}
The Lagrange function  corresponding to the BGPP metric 
reads
\beq
L=\frac 12\Big(f(t)^2\,\dot{t}^2+a(t)^2\,s_1^2+b(t)^2\,s_2^2+c(t)^2\,s_3^2\Big),\eeq
where 
\beq\barr{l}
s_1=-\sin\psi\,\dot{\tht}+\sin\tht\,\cos\psi\,\dot{\phi}, \\[4mm]
s_2=+\cos\psi\,\dot{\tht}+\sin\tht\,\sin\psi\,\dot{\phi}, \\[4mm]
s_3=\dot{\psi}+\cos\tht\,\dot{\phi}, \earr
\eeq
and  the dot over a symbol denotes the derivation with respect to  parameter $\lambda$ along geodesic, while the prime denotes the derivation with respect to coordinate $t$.  The canonical  momenta conjugated to coordinates  $t,\theta,\phi,\psi$  are
 the following:
\beq\barr{l}\dst 
P_t=\frac{\pt L}{\pt\dot{t}}=f(t)^2\,\dot{t},\qq 
P_{\tht}=\frac{\pt L}{\pt\dot{\tht}}=-a(t)^2\,s_1\,\sin\psi+b(t)^2\,s_2\,\cos\psi,\\[4mm]\dst 
P_{\phi}=\frac{\pt L}{\pt\dot{\phi}}=
a(t)^2\,s_1\,\sin\tht\,\cos\psi+b(t)^2\,s_2\,\sin\tht\,\sin\psi+c(t)^2\,s_3\,\cos\tht,\,\,\,
P_{\psi}=\frac{\pt L}{\pt\dot{\psi}}=c(t)^2\,s_3,
\earr\eeq
and Hamilton function takes the form
\begin{equation}
\begin{split}
H&=\frac{\left(\csc \theta  \cos \psi(P_\phi-P_\psi \cos
   \theta )-P_\theta
   \sin \psi \right)^2}{2
   a(t)^2}+\frac{\left(P_\theta
   \cos \psi +\sin \psi 
   (P_\phi \csc\theta
   -P_\psi \cot \theta)\right)^2}{2
   b(t)^2}\\
   &+\frac{P_\psi^2}{2 c(t)^2}+\frac{P_t^2}{2 f(t)^2}.
   \end{split}
   \label{eq:hamiltwith4deg}
\end{equation}
The corresponding Hamilton equations are quite lengthy   but one can simplify them introducing the following variables
\beq\barr{l}\dst 
M_1=-\sin\psi\,P_{\tht}+\frac{\cos\psi}{\sin\tht}\,P_{\phi}
-\frac{\cos\psi}{\tan\tht}\,P_{\psi}=a(t)^2\,s_1,\\[4mm]\dst 
M_2=+\cos\psi\,P_{\tht}+\frac{\sin\psi}{\sin\tht}\,P_{\phi}
-\frac{\sin\psi}{\tan\tht}\,P_{\psi}=b(t)^2\,s_2,\\[4mm] M_3=P_{\psi}=c(t)^2\,s_3.\earr
\label{eq:mmm}
\eeq
The Poisson brackets of these functions are
\begin{equation}
   \{M_i,M_j\}=\eps_{ijk}\,M_k.   
\end{equation}
In the non-canonical variables $(t, P_t, M_1, M_2, M_3, \phi,\theta,\psi)$ 
the Hamiltonian~\eqref{eq:hamiltwith4deg} has particularly simple form 
\beq
\label{Ham}
H=\frac 12\left(\frac{P_t^2}{f(t)^2}+\frac{M_1^2}{a(t)^2}+\frac{M_2^2}{b(t)^2}+\frac{M_3^2}{c(t)^2}\right).
\eeq 
However, the symplectic form is no longer constant. In these non-canonical variables  the $8\times 8$ Poisson tensor reads 
\begin{equation}
\label{eq:P8}
 J  =  \begin{bmatrix}
    J_2   & 0  & 0 \\
    0     & \widehat M & -K^T \\
    0   &  K &  0
 \end{bmatrix},
\end{equation}
where 
\begin{equation}
   \label{eq:mtrx}
   \widehat M  = 
   \begin{bmatrix}
   0&M_3&-M_2\\
-M_3&0&M_1\\
   M_2&-M_1&0
   \end{bmatrix}, \,\,
   K = \begin{bmatrix}
      \csc\theta \cos \psi  & \csc\theta \sin \psi & 0 \\
      -\sin \psi  & \cos \psi &  0\\
      -\cot\theta \cos \psi  & -\cot\theta \sin \psi & 1
   \end{bmatrix},\,\, J_2=\begin{bmatrix}
       0&1\\
       -1&0
   \end{bmatrix}
\end{equation}
and the explicit form of Hamilton equations corresponding to this structure is the following
\begin{equation}
    \begin{split}
       \dot{t}=&\frac{1}{f(t)^2}P_t,  \\ 
       \dot{P}_t=&\frac{f'(t)}{f^3(t)}P_t^2+\frac{a'(t)}{a^3(t)}M_1^2+\frac{b'(t)}{b^3(t)}M_2^2+\frac{c'(t)}{c^3(t)}M_3^2,\\
       \dot{M}_1=&\left(\frac 1{b(t)^2}-\frac 1{c(t)^2}\right)M_2\,M_3=\frac{(t_3-t_2)}{A(t)B(t)C(t)}M_2\,M_3,\\
\dot{M}_2=&\left(\frac 1{c(t)^2}-\frac 1{a(t)^2}\right)M_3\,M_1=\frac{(t_1-t_3)}{A(t)B(t)C(t)}M_3\,M_1,\\
\dot{M}_3=&\left(\frac 1{a(t)^2}-\frac 1{b(t)^2}\right)M_1\,M_2=\frac{(t_2-t_1)}{A(t)B(t)C(t)}M_1\,M_2, \\
  \dot{\phi}=&\csc \theta
   \left(\frac{M_1 \cos\psi}{a(t)^2}+\frac{M_2
   \sin\psi }{b(t)^2}\right),\\
   \dot{\theta}=&\frac{M_2 \cos \psi}{b(t)^2}-\frac{M_1 \sin
   \psi }{a(t)^2},\\
    \dot{\psi}=&-\frac{M_1 \cot \theta 
   \cos \psi}{a(t)^2}-\frac{M_2 \cot\theta  \sin \psi}{b(t)^2}+\frac{M_3}{c(t)^2}.
    \end{split}
    \label{eq:systemmix}
\end{equation}

This system is Hamiltonian with four degrees of freedom and for its integrability four independent  commuting first integrals are necessary

\begin{nth} The geodesic flow \eqref{eq:systemmix} of the BGPP Hamiltonian  is integrable in Liouville sense.\end{nth}

\nin{\bf Proof:} Apart from the Hamiltonian we have to exhibit 3 integrals in involution and algebraically independent. The first one follows from the cyclic variable $\phi$ and  it is the corresponding momentum $P_{\phi}$ and additionally exist two others which have the following forms
\begin{equation}
\begin{split}
&P_{\phi}=\sin \theta  (M_1 \cos \psi +M_2 \sin \psi )+M_3 \cos \theta ,\\
&{\cal C}=M_1^2+M_2^2+M_3^2 ,\\
&{\cal I}=t_1M_1^2+t_2M_2^2+t_3M_3^2,
   \end{split}
   \label{eq:cycalki}
\end{equation}
and  $H$,  $P_\phi$, ${\cal C}$ and ${\cal I}$ are independent and pairwise commuting with the standard Poisson bracket.
$\quad\Box$

\section{Reduced system of the geodesic flow}
\label{sec:reduced_system}
Let us notice that system \eqref{eq:systemmix} contains a closed subsystem 
\beq\barr{ll}\dst 
 \qq\dot{t}=\frac{1}{f(t)^2}P_t,\\[4mm]\dst 
\qq\dot{P}_t=\frac{f'(t)}{f^3(t)}P_t^2+\frac{a'(t)}{a^3(t)}M_1^2+\frac{b'(t)}{b^3(t)}M_2^2+\frac{c'(t)}{c^3(t)}M_3^2,\\[4mm]\dst 
\qq\dot{M}_1=\left(\frac 1{b(t)^2}-\frac 1{c(t)^2}\right)M_2\,M_3=4f(t)^2(t_3-t_2)M_2\,M_3,\\[4mm]\dst 
\qq\dot{M}_2=\left(\frac 1{c(t)^2}-\frac 1{a(t)^2}\right)M_3\,M_1=4f(t)^2(t_1-t_3)M_3\,M_1,\\[4mm]\dst
\qq\dot{M}_3=\left(\frac 1{a(t)^2}-\frac 1{b(t)^2}\right)M_1\,M_2=4f(t)^2(t_2-t_1)M_1\,M_2.\earr
\label{eq:system}
\eeq
These equations are of the form 
\begin{equation}
\frac{\mathrm{d}X_i}{\mathrm{d}\lambda}=\sum_{j=1}^5P_{ij}\frac{\partial H}{\partial X_j},
\label{eq:Poissonsys}
\end{equation}
where  $X=[t,P_t,M_1,M_2,M_3]^T$, function $H$ is given by \eqref{Ham}, and 
\begin{equation}
P=[P_{ij}]=
\begin{bmatrix}
   J_2   & 0  \\
   0     & \widehat M \\
\end{bmatrix}.
\label{eq: Poissontens}
\end{equation}
 So, they are Hamiltonian with Poisson structure given by $P$. This Poisson tensor defines the Poisson bracket. Namely, for any two functions $F(X)$ and $G(X)$ we set
\begin{equation}
\{F,G\}(X)=\frac{\partial F^T}{\partial X}P\frac{\partial G}{\partial X}.
\label{eq:Poisson}
\end{equation}
Clearly, bracket $\{\cdot,\cdot\}$ is antisymmetric and satisfies the Leibniz rule and the Jacobi identity.  
However, it is degenerated.
The Casimir function 
\beq
{\cal C}=M_1^2+M_2^2+M_3^2, 
\label{eq:casim}
\eeq
commute with any function $F(X)$: $\{{\cal C},F\}(X)=0$. On a symplectic leaf
${\cal C}(X)= \alpha>0$ this system becomes Hamiltonian with two degrees of
freedom, thus for its integrability one more functionally independent first
integrals is necessary. This reduction  from four to two degrees of freedom is
possible thanks to the $\mathrm{SO}(3,\mathbb{R})$ symmetry of the original
system. 

We have the following. 
\begin{nth} The reduced system corresponding to the geodesic flow of the BGPP Hamiltonian is integrable in Liouville sense on its symplectic leaf.\end{nth}
\nin{\bf Proof:} 
Apart from the Casimir function (\ref{eq:casim}) and Hamiltonian (\ref{Ham})
Poisson system (\ref{eq:system}) possesses  additional first integral ${\cal I}$
defined in \eqref{eq:cycalki}. Functions ${\cal C}$, $H$ and ${\cal I}$ are
functionally independent and their Poisson brackets (\ref{eq:Poisson}) vanish.
Moreover, on a level ${\cal C}(X)= \alpha>0$ the differentials $d H(X)$ are $ d
{\cal I}(X)$ are linearly independent. Hence, restriction of $H$ and ${\cal I}$
to this level gives a Liouville integrable system. $\quad\Box$ 

\section{Integration of the reduced system}
\label{sec:reducedintegration}

If the system is integrable, then one would like to give on explicit formulae for its solutions. However, generally it is a highly non-trivial task. In particular, it is true for the considered system. 

At first, let us notice that the last three equations in the reduced  system~\eqref{eq:system} have the form
of the Euler  equations of a rigid body whose moments of inertia are
controlled by  dynamical variables $(t,P_t)$.  Our idea how to effectively
perform the integration is similar to that applied in integration of the Kepler
problem: on a given fixed level of first integrals we introduce implicitly new
independent variable $\tau$, then we give solutions as function of $\tau$. 
Clearly the last step of this approach has to be a formula 
 that defines implicitly or explicitly  $\tau(\lambda)$. 
In this section  we  assume that  $0\leq t_1<t_2<t_3$.

On the common level sets of the first integrals ${\cal C}=m^2$, ${\cal I}=n^2$
and $H=e$ we have

\[\left(\frac{dt}{d\la}\right)^2=\frac{P_t^2}{f(t)^4}=\frac{2e}{f(t)^2}-4(A(t)^2\,M_1^2+B(t)^2\,M_2^2+C(t)^2\,M_3^2)=\frac{2e}{f(t)^2}-4m^2\,t+4n^2,\]
leading to
\beq\label{sep}
\left(\frac{dt}{d\lambda}\right)^2={S}(t)\qq\qq { S}(t)=4\left(n^2-m^2t+2e\sqrt{(t-t_1)(t-t_2)(t-t_3)}\right),
\eeq
that, in principle, enables to express $t$ as a function of $\lambda$. However,
we do not know if it is possible to find it in an analytic way. In fact,
notice that the genus of the algebraic curve $y^2={S}(t)$ is generically equal
to 3.

If we introduce the new independent variable $\tau$ such that
\begin{equation}
\frac{\mathrm{d}\tau}{\mathrm{d}\lambda}=4f(t)^2,
\label{eq:changetotau}
\end{equation}
 then the last three equations of   system \eqref{eq:system} can be written as
\begin{equation}
\frac{dM_1}{d\tau}=(t_3-t_2)M_2\,M_3, \qq  
\frac{dM_2}{d\tau}=(t_1-t_3)M_3\,M_1, \qq 
\frac{dM_3}{d\tau}=(t_2-t_1)M_1\,M_2,
\label{eq:Euler}
\end{equation}
in which we can recognise Euler equations for a rigid body. 

Notice that equations  \eqref{eq:changetotau} and \eqref{sep} imply
\begin{equation}
\frac{\mathrm{d}\tau}{\mathrm{d}t} =\frac{4f(t)^2}{\frac{\mathrm{d}t}{\mathrm{d}\lambda}}  =\frac{4f(t)^2}{\sqrt{{ S}(t)} },
\end{equation}
that gives function $\tau=\tau(t)$  as a  quadrature
\begin{equation}
\tau(t)=\int_{t_3}^t\,\frac{du}{\sqrt{(u-t_1)(u-t_2)(u-t_3)\,{S}(u)}}.
\label{eq:taut}
\end{equation}
The main difference between the Euler equations and  equations \eqref{eq:Euler}
is that time is replaced by the variable $\tau$, given by the Abelian integral
\eqref{eq:taut}.
 
Following calculations of explicit solutions for an asymmetric top given e.g. in
\cite[Chap. II, Sec. 102]{Markeev} or  \cite[par. 37]{Landau} on a common level
of first integrals  ${\cal C}=m^2$, ${\cal I}=n^2$ depending on values of $n^2$
and $m^2$ explicit solutions are following:\\ 
in case I when $t_3> \tfrac{n^2}{m^2}>t_2>t_1$:
\[
\begin{split}
&M_1=\mp\sqrt{\frac{m^2t_3-n^2}{t_3-t_1}}\,\cn(\sigma,k),\,\, M_2=\pm\sqrt{\frac{m^2t_3-n^2}{t_3-t_2}}\,\sn(\sigma,k),\,\,
M_3=\sqrt{\frac{n^2-m^2t_1}{t_3-t_1}}\,\dn(\sigma,k),\\
&\sigma=\sqrt{(t_3-t_2)(n^2-m^2t_1)}\,\tau, \quad k^2=\frac{(t_2-t_1)}{(t_3-t_2)}\frac{(m^2t_3-n^2)}{(n^2-m^2t_1)}<1;
\end{split}
\]
in case II when $t_2> \tfrac{n^2}{m^2}>t_1$:
\[
\begin{split}
&M_1=\sqrt{\frac{m^2t_3-n^2}{t_3-t_1}}\,\dn(\sigma,k),\,\, M_2=\pm\sqrt{\frac{n^2-m^2t_1}{t_2-t_1}}\,\sn(\sigma,k),\,\,
M_3=\mp\sqrt{\frac{n^2-m^2t_1}{t_3-t_1}}\,\cn(\sigma,k),\\
&\sigma=\sqrt{(t_2-t_1)(m^2t_3-n^2)}\,\tau,\quad
k^2=\frac{(t_3-t_2)}{(t_2-t_1)}\frac{(n^2-m^2t_1)}{(m^2t_3-n^2)}<1;
\end{split}
\]
in case III when  equality $\tfrac{n^2}{m^2}=t_2$ holds
\[
\begin{split}
&M_1=m\sqrt{\frac{t_3-t_2}{t_3-t_1}}\frac{1}{\cosh\sigma},\quad M_2=m\tanh\sigma,\quad M_3=-m\sqrt{\frac{t_2-t_1}{t_3-t_1}}\frac{1}{\cosh\sigma},\\
& \sigma=m\sqrt{(t_3-t_2)(t_2-t_1)}\,\tau.
\end{split}
\]

Let us notice that $\tau$ is expressible by  quadrature of $t$ given in
\eqref{eq:taut} thus solutions are no longer double-periodic functions of $t$ as
it was for a free rigid body.

Here we remark that our solution is nor fully satisfactory. We did not express analytically $t$ as a function of $\tau$ and because of this we did not express $\tau$ as a function of $\lambda$. 

\section{Eguchi-Hanson limit}
\label{sec:degenerated_system}

In the previous section we considered generic case when constants
$(t_1,t_2,t_3)$ are pairwise different. If $t_2=t_3$, or $t_1=t_2$, the either
$M_1$, or $M_3 $ is a first integral.   Interpreting equations for $M_i$ as Euler
equations, these cases correspond to an axially symmetric body. In both cases
$M_i(\tau)$ are  given by trigonometric functions. In the paper \cite{BGPP} the
metrics corresponding these limiting cases are called the  Eguchi-Hanson metric
I and metric II, respectivly.  Both of them are discussed in the second part of
the paper. See also \cite{EH} where these metric have been introduced. As it was
pointed in these references the first metric is not complete. So, here we will
focus on  the second case only, and we will call it the Eguchi-Hanson metric, or
the Eguchi-Hanson limit.

It is convenient to assume here that $t_3<t_2<t_1<t$. Then the Eguchi-Hanson
limit we are going to investigate is given by $t_1=t_2$. After the change of
variable $t\, \to \,\rho=\sqrt{t-t_3}$ the BGPP metric \eqref{met0}  becomes
\beq
g=\frac{\rho\,d\rho^2}{\rho^2-\ga^2}+\rho(\si_1^2+\si_2^2)+\frac{\rho^2-\ga^2}{\rho}\si_3^2,\qq \ga^2=t_1-t_3.
\eeq
Hence, we have $\rho> \ga$ and in this expression we recognize the Eguchi-Hanson metric derived in \cite{EH} 
\[
g_{\mathrm{II}}=\frac{4r^4}{r^4-a^4}dr^2+r^2(\sigma_1^2+\sigma_2)^2+\frac{r^4-a^4}{r^2}\sigma_3^2,
\]
when we substitute $r^2=\rho$ and $a^4=\gamma^2$. Despite the apparent
singularity $\rho=\ga$ (a ``bolt" according to Gibbons and Hawking) let us
mention that this metric is complete and defined on  the cotangent bundle of the
two-sphere \cite{EH1}. Its geodesic flow at
the classical and quantum levels was first discussed in \cite{Mi}. However, the
author, precluding elliptic integrals, did not give the explicit flow formulae
in general. These can be conveniently derived within our approach.

The Hamiltonian \eqref{Ham} now reads
\beq
H=\frac 12\left(\frac{\rho^2-\ga^2}{\rho}\,P_{\rho}^2+\frac{M_1^2+M_2^2}{\rho}+\frac{\rho}{\rho^2-\ga^2}M_3^2\right),
\eeq
and equations \eqref{eq:system} transforms into
\beq\barr{l}\dst
\dot{\rho}=\frac{(\rho^2-\ga^2)P_{\rho}}{\rho},\\
\dst \dot{P}_{\rho}=\frac 12\left(-\frac{P_{\rho}^2(\rho^2+\ga^2)}{\rho^2}+\frac{M_1^2+M_2^2}{\rho^2}+\frac{M_3^2(\rho^2+\ga^2)}{(\rho^2-\ga^2)^2}\right),\\
\dst 
\dot{M}_1=-\frac{\ga^2\,M_2M_3}{\rho(\rho^2-\ga^2)},\\
\dst 
\dot{M}_2=+\frac{\ga^2\,M_3M_1}{\rho(\rho^2-\ga^2)},\\
\dst 
\dot{M}_3=0.\earr
\eeq
This system has first integrals 
\[H, \qq M_3,\qq M_1^2+M_2^2,\]
 i.e. one of them is linear, and two others are quadratic in variables. For the BGPP metric the reduced system has three  first integrals quadratic in variables.

On a common set of first integrals $M_3=m_3$, $M_1^2+M_2^2=\mu^2$ and $H=e$ we have
\beq
\label{r2dl}
\left(\frac{d\rho}{d\la}\right)^2=\frac{R(\rho)}{\rho^2}, \quad \mbox{where} \quad R(\rho)=2e\rho^3-(\mu^2+m_3^2)\rho^2-2e\ga^2\rho+\ga^2\mu^2.
\eeq
Let us notice that the discriminant of the polynomial $R(\rho)$ is always positive
\beq
\label{dis}
{\rm disc}\,R(\rho)=4\ga^2\Big[(\mu^4-4\ga^2e^2)^2+m_3^2\Big(\ga^2e^2(20\mu^2+m_3^2)+\mu^2(3\mu^4+m_3^4+3\mu^2m_3^2)\Big)\Big],
\eeq
except for one case which will be considered later. Thus we have three real roots $\rho_1<\rho_2<\rho_3$ and we put $R(\rho)=2e(\rho-\rho_1)(\rho-\rho_2)(\rho-\rho_3)$.
Since $R(\pm\ga)<0$ and $R(0)>0$ we conclude to
\[ -\ga<\rho_1<0<\rho_2<\ga<\rho_3.\]
Notice that equation \eqref{r2dl} defines domains of possible motion. Curve $y^2=R(\rho)/\rho^2$ in half plane has two connected components, see Fig. 1. 
\begin{figure}[h!tp]
   \centering
   {\includegraphics[width=.47\linewidth]{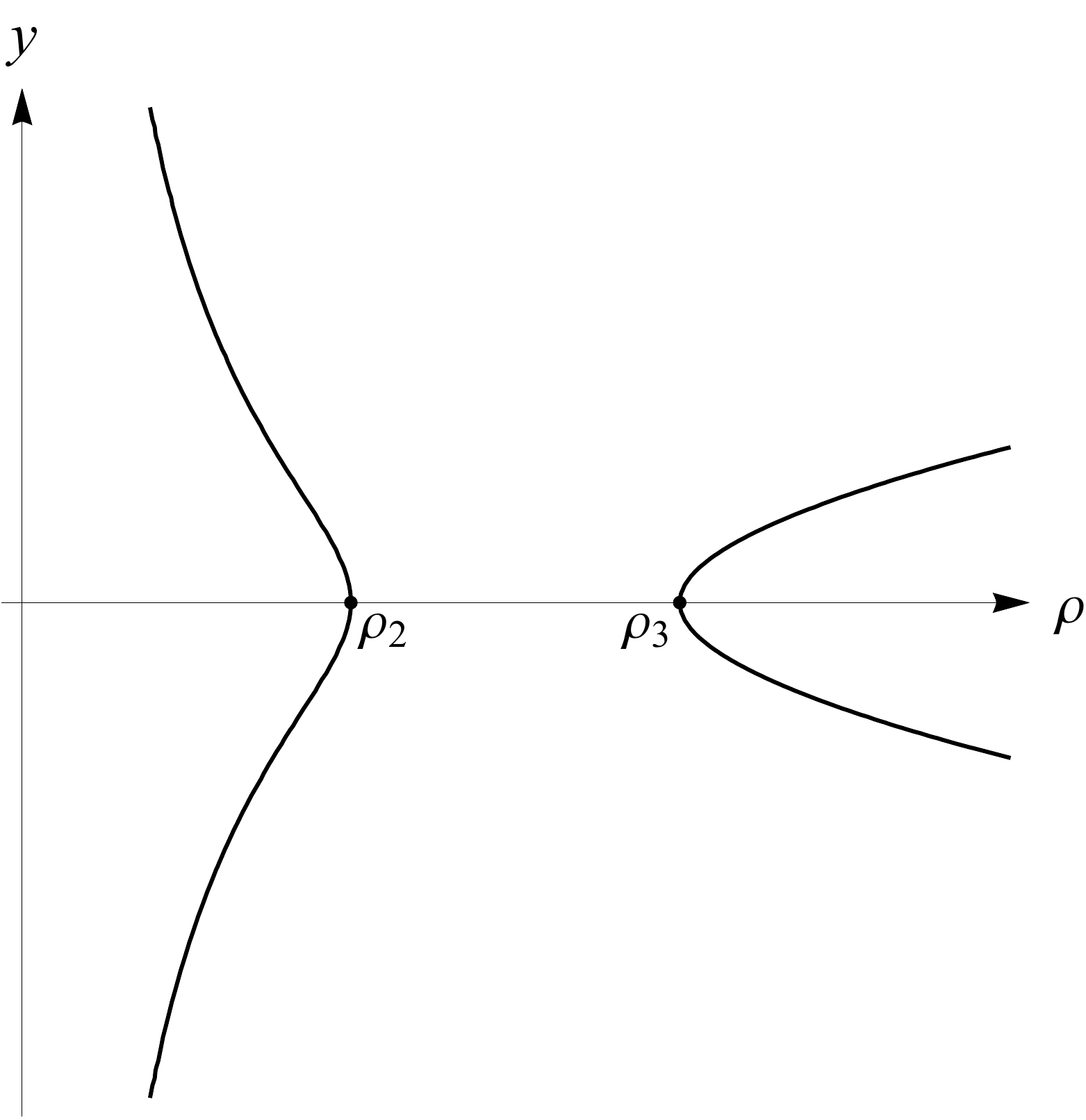}}
   \caption{Phase curves $y^2=R(\rho)/\rho^2$ \label{curvesy}}
 \end{figure}
For the first component $0<\rho\leq\rho_1$ and for the second $\rho \geq \rho_3$. Since $\rho>\ga$ we have to consider only the case $\rho\geq\rho_3$.

Now we introduce the variable $\tau$ such that
\beq
\label{dtaudl}
\frac{d\tau}{d\la}=\frac 1{\rho(\rho^2-\ga^2)},
\eeq
which gives
\beq
\frac{dM_1}{d\tau}=-\om\,M_2,\qq\qq \frac{dM_2}{d\tau}=\om\,M_1,\qq\qq \om=\ga^2\,m_3,
\eeq
and is easily integrated to
\[
M_1=\mu\sin(\omega\tau+\varphi_0),\quad M_2=-\mu\cos(\omega\tau+\varphi_0),
\]
where $\mu$ and $\varphi_0$ are integration constants.

Equations \eqref{r2dl} and \eqref{dtaudl} imply that
\beq
\left(\frac{d\rho}{d\tau}\right)^2=(\rho^2-\ga^2)^2R(\rho).
\eeq
Integrating
\beq
d\tau=\frac{d\rho}{(\rho^2-\ga^2)\sqrt{R(\rho)}}
\eeq
with assumption $\rho\geq\rho_3$ and using formulae 3.137
no 8 from [10] we obtain
\beq
\tau=\frac 1{\ga\sqrt{2e}\sqrt{\rho_3-\rho_1}(\ga^2-\rho_1^2)}\Big[2\ga\,F(\si,k)-(\ga+\rho_1)\Pi(\si,n_-,k)-(\ga-\rho_1)\Pi(\si,n_+,k)\Big],
\eeq
where $F(z,k)$ and $\Pi(z,n,k)$ are the incomplete elliptic integrals of the first and the third kind defined in 8.111 no 2-4 from [10]. In the above formula
\[ \si=\arcsin\sqrt{\frac{\rho_3-\rho_1}{\rho-\rho_1}}, \qq n_{\pm}=\frac{\rho_1\pm\ga}{\rho_1-\rho_3}, \qq k=\sqrt{\frac{\rho_2-\rho_1}{\rho_3-\rho_1}}.\]
Discriminant \eqref{dis} is positive except for $m_3=0$ and $\mu^2=2e\ga$ in which case $R(\rho)$ has a double root
\[  R(\rho)=(\rho-\ga)^2(\rho+\ga).\]
For this case we obtain
\beq
\tau=\frac 1{4\sqrt{2e}\ga^2}\left[\frac{(\ga-3\rho)}{(\rho-\ga)\sqrt{\rho+\ga}}+\frac{3}{2\sqrt{2\ga}}\ln\left(\frac{\sqrt{\rho+\ga}+\sqrt{2\ga}}{\sqrt{\rho+\ga}-\sqrt{2\ga}}\right)\right],\qq \rho>\ga.
\eeq

\section{Conclusion}
\label{sec:conclusion}
In this article we analyze the integrability of the geodesic flow corresponding
to the BGPP metric in generic and non-generic degenerated cases. In all these
cases the corresponding Hamiltonian systems are integrable in the Liouville
sense. The $\mathrm{SO}(3,\mathbb{R})$ symmetry of the model allows to reduce
the system to two degrees of freedom. After introducing variables $M_1,M_2,M_3$
by formulae  \eqref{eq:mmm} obtained system splits into two subsystems and the
reduced system also appear to be integrable in all cases.  Moreover, the reduced
system can be solved explicitly as functions of a new variable $\tau$ that is
related to $t$ by a  quadrature. When the BGPP metric reduces to the
Eguchi-Hanson metric it is possible to calculate explicitly the mentioned
quadrature. An important and interesting question remains open: is the geodesic
flow of the A-H metric integrable? 

\vspace{3mm}{\bf Acknowledgments:} We are greatly indebted to Professor S. Shevchishin for the determination of the genus of the algebraic curve $y^2= S(t)$ in Section~\ref{sec:reducedintegration}. 

This research has been partially founded by The
National Science Center of Poland under Grant No.
2020/39/D/ST1/01632. For the purpose of Open
Access, the authors have applied a CC-BY public
copyright license to any Author Accepted Manuscript
(AAM) version arising from this submission.



\begin{thebibliography}{3333}
\bibitem{AH1} M. F. Atiyah and N. J. Hitchin, {\sl Phys. Lett.}, {\bf A 107} (1985) 21-25.
\bibitem{AH2}  M. F. Atiyah and N. J. Hitchin, {\em The Geometry and Dynamics of Magnetic Monopoles}, Princeton University Press (2014).
\bibitem{Audin} M. Audin, {\em Spinning tops}, Cambridge Studies in Advanced Mathematics 51, Cambridge University Press, Cambridge (1996).
\bibitem{BM} L. Bates and R. Montgomery, {\sl Commun. Math. Phys.}, {\bf 118} (1988) 635-640.
\bibitem{BGPP} V. Belinskii, G. W. Gibbons, D. N. Page and C. N. Pope, {\sl Phys. Lett.}, {\bf B 76} (1978) 433-435.
\bibitem{Do} D. Dold, {\sl Class. Quantum Grav.}, {\bf 35}(9), (2018) 095012.
\bibitem{Gh} A. M. Ghezelbash, {\sl Class. Quantum Grav.}, {\bf 39} (7), (2022) 075012.
\bibitem{GR} G. W. Gibbons and P. J. Ruback, {\sl Comm. Math. Phys.}, {\bf 115} (1968) 267-300.
\bibitem{GrR} I. S Gradshteyn and I. M. Ryzhik,  {\em  Table of integrals, series, and products},   Eighth edition,  Elsevier/Academic Press, Amsterdam (2015).
\bibitem{EH} T. Eguchi and A. J. Hanson, {\sl Phys. Lett.}, {\bf 74B} (1978) 249-251.
\bibitem{EH1} T. Eguchi and A. J. Hanson, {\sl Ann. Phys.},  {\bf 120} (1979) 82-106.
\bibitem{GM} G. W. Gibbons and N. S. Manton, {\sl Nucl. Phys.} {\bf B274} (1986) 183-224.
\bibitem{GOVR} G. W. Gibbons, D. Olivier, G. Valent and P. J. Ruback, {\sl Nucl. Phys.} {\bf B 296} (1988) 676-696.
\bibitem{Gi} G. W. Gibbons, {\sl Class. Quantum Grav.}, {\bf 20} (2003) 4401-4408.
\bibitem{Hi} N. Hitchin, {\sl NATO Advanced Study Institute}, Les Presses de l'Universit\'e de Montr\'eal, Montr\'eal (Qu\'ebec) (1987).
\bibitem{KL} D. Katona and J. Lucietti, {\sl Commun. Math. Phys.}, {\bf 399} (2023)
\bibitem{Landau} L.\,D. Landau and E.\,M. Lifshitz, {\em Mechanics},Vol. 1 of Course of Theoretical Physics, Third Edition, Butterworth-Heinenann, Oxford (1997).
\bibitem{Markeev} A.\,P. Markeev, {\em Teoreticheskaya mekhanika}, 2nd ed.,  R{\&}C Dynamics, Institute of Computer Science, Izhevsk (1999), (In Russian).
\bibitem{Mi} S. Mignemi, {\sl J. Math. Phys.}, {\bf 32} (1991) 3047-3054.
\bibitem{Va1} G. Valent, {\sl Comm. Math. Phys.}, {\bf 244} (2004) 571-594.
\bibitem{Va2} G. Valent and H. Ben Yahia, {\sl Class. Quantum Grav.}, {\bf 24} (2007) 255-276.
\bibitem{Wo} M. P. Wojtkowski, {\sl Bull. Amer. Math. Soc.}, {\bf 18} (1988) 179-183.
\end{thebibliography}
\end{document}